\documentclass[aps,prd,onecolumn,showpacs,showkeys,amsmath,amssymb]{revtex4}
\usepackage{amsfonts}
\usepackage{graphicx}
\usepackage{dcolumn}
\usepackage{bm}
\usepackage[dvips]{color}
\begin{document}
%
%
\title{The Possible $J^{PC} = 0^{+-}$ Exotic State}
\author{Meng-Lin Du}
\email{du_menglin@pku.edu.cn}
\author{Wei Chen}
\email{boya@pku.edu.cn}

\author{Xiao-Lin Chen}
\email{chenxl@pku.edu.cn}\affiliation{Department of Physics
and State Key Laboratory of Nuclear Physics and Technology\\
Peking University, Beijing 100871, China  }
\author{Shi-Lin Zhu}
\email{zhusl@pku.edu.cn} \affiliation{Department of Physics
and State Key Laboratory of Nuclear Physics and Technology\\
and Center of High Energy Physics, Peking University, Beijing
100871, China }
\begin{abstract}

We study the possible exotic states with $J^{PC} = 0^{+-}$ using
the tetraquark interpolating currents with the QCD sum rule
approach. The extracted masses are around 4.85 GeV for the
charmonium-like states and 11.25 GeV for the bottomomium-like
states. There is no working region for the light tetraquark
currents, which implies the light $0^{+-}$ state may not exist
below 2 GeV.

\end{abstract}
\maketitle
\pagenumbering{arabic}
%
%
\section{Introduction}\label{sec:intro}
%
Up to now most of the hadrons observed experimetally can be
interpreted as the $q \bar q$/$qqq$ states in the quark
model~\cite{Amsler:2008zzb,Klempt:2007cp}. However there has
accumulated some evidence of the exotic state with $J^{PC} =
1^{-+}$ ~\cite{Adams:2006sa,Abele:1999tf,Thompson:1997bs}. Such a
quantum number is not accessible for a conventional meson composed
of a pair of quark and anti-quark in the non-relativistic quark
model. Sometimes these states are named as exotic states although
all the $J^{PC}$ quantum numbers are allowed in QCD.

For a neutral quark model $q\bar{q}$ state, $J=0$ ensures $L=S$
hence $C=(-)^{L+S}=+1$. Therefore, there exist two possible exotic
states with $J^{PC}=0^{--}$ and $0^{+-}$. It's also interesting to
note that the $J^{PC}$ quantum number of the local operators
composed of a pair of the gluon field strength tensor is either
$0^{++}$ or $0^{-+}$.

On the other hand, the tetraquark operators may carry the $0^{--}$
and $0^{+-}$ quantum numbers. In fact, the $0^{--}$ state was
investigated systematically using the tetraquark currents with the
QCD sum rule method ~\cite{Jiao:2009,Wei:2010}. As a byproduct, it
was noted that there does not exist any tetraquark interpolating
current without derivative for the $J^{PC}=0^{+-}$ case.

With the similar formalism, one may construct the possible
$0^{+-}$ tetraquark current by introducing derivatives. There are
two kinds of constructions either with the $qq$ basis or with the
$\bar qq$ basis: $(qq)(\bar{q}\bar{q})$ and
$(\bar{q}q)(\bar{q}q)$. However, they can be related to each other
by the Fierz transformation~\cite{Jiao:2009}. In this work we use
the first set. With these independent $0^{+-}$ currents, we
perform the QCD sum rule analysis and extract the masses of the
corresponding currents.

This paper is organized as follows. In Sec.~\ref{sec:current}, we
construct the tetraquark currents with $J^{PC} = 0^{+-}$ using the
diquark ($qq$) and antidiquark ($\bar q \bar q$) fields. In
Sec.~\ref{sec:ope}, we calculate the correlation functions and
spectral densities of the interpolating currents and collect them
in the Appendix~\ref{sec:rhos}. We perform the numerical analysis and extract the
masses in Sec.~\ref{sec:num} for the light and heavy systems
respectively. The last section is a brief summary.

%
%
\section{tetraquark interpolating currents}\label{sec:current}
%
It was shown that the $J^{PC}=0^{+-}$ tetraquark interpolating
currents without derivatives do not exist \cite{Jiao:2009}. So in
this work we construct the $0^{+-}$ currents with the derivatives
following the similar steps as in Ref. \cite{Jiao:2009}. We first
construct two independent tetraquark fields:
\begin{eqnarray}\label{no1}
&&A_{abcd}=(q_{1a}^TC\gamma^{\mu}q_{2b})(\bar{q}_{3c}\stackrel{\leftrightarrow}{D}_\mu C\bar{q}^T_{4d}) \, ,\\
\nonumber &&
P_{abcd}=(q_{1a}^TC\gamma^{\mu}\gamma_5q_{2b})(\bar{q}_{3c}\stackrel{\leftrightarrow}{D}_\mu
\gamma_5C\bar{q}^T_{4d}) .
\end{eqnarray}
where $q_{1-4}$ represents the flavor of quarks, and $a-d$ stands
for the color indices,
$\stackrel{\leftrightarrow}{D}_\mu={\overrightarrow
D}_\mu-{\overleftarrow D}_\mu$, ${\overrightarrow
D}_\mu={\overrightarrow
\partial}_\mu+igA_\mu^at^a$.
It is understood that the index $c$ is the color index of
$\left(\bar q {\overleftarrow D}_\mu\right)_c$. In Eqs.(\ref{no1}) and (\ref{equ:current}) we have used the shorthand
notation to simply the expression.

To compose the color singlet tetraquark currents, the diquark and
antidiquark should have the same color and spin symmetries.
Therefore the color structure of the tetraquark is either $\mathbf
6 \otimes \mathbf { \bar 6}$ or $\mathbf { \bar 3 }\otimes \mathbf
3$, which is denoted by labels  $\mathbf  6 $ and $\mathbf 3 $
respectively. Details can be found in Ref.~\cite{Jiao:2009}.
Considering both the color and Lorentz structures, we can obtain
the currents with $J^{PC}=0^{+-}$:
\begin{equation}
\begin{split}
&\eta_1(x)=q_{1a}^TC\gamma^\mu q_{2b}(\bar{q}_{1a}\stackrel{\leftrightarrow}{D}_\mu C\bar{q}_{2b}^T+\bar{q}_{1b}\stackrel{\leftrightarrow}{D}_\mu C\bar{q}^T_{2a})-q_{1a}^TC\stackrel{\leftrightarrow}{D}_\mu q_{2b}(\bar{q}_{1a}\gamma^\mu C\bar{q}_{2b}^T+\bar{q}_{1b} \gamma^\mu C\bar{q}^T_{2a}),\\
&\eta_2(x)=q_{1a}^TC\gamma^\mu q_{2b}(\bar{q}_{1a}\stackrel{\leftrightarrow}{D}_\mu C\bar{q}_{2b}^T-\bar{q}_{1b}\stackrel{\leftrightarrow}{D}_\mu C\bar{q}^T_{2a})-q_{1a}^TC\stackrel{\leftrightarrow}{D}_\mu q_{2b}(\bar{q}_{1a}\gamma^\mu C\bar{q}_{2b}^T-\bar{q}_{1b} \gamma^\mu C\bar{q}^T_{2a}),\\
&\eta_3(x)=q_{1a}^TC\gamma^\mu \gamma_5q_{2b}(\bar{q}_{1a}\stackrel{\leftrightarrow}{D}_\mu \gamma_5C\bar{q}_{2b}^T+\bar{q}_{1b}\stackrel{\leftrightarrow}{D}_\mu \gamma_5C\bar{q}^T_{2a})-q_{1a}^TC\stackrel{\leftrightarrow}{D}_\mu\gamma_5 q_{2b}(\bar{q}_{1a}\gamma^\mu \gamma_5C\bar{q}_{2b}^T+\bar{q}_{1b} \gamma^\mu \gamma_5C\bar{q}^T_{2a}),\\
&\eta_4(x)=q_{1a}^TC\gamma^\mu
\gamma_5q_{2b}(\bar{q}_{1a}\stackrel{\leftrightarrow}{D}_\mu
\gamma_5C\bar{q}_{2b}^T-\bar{q}_{1b}\stackrel{\leftrightarrow}{D}_\mu
\gamma_5C\bar{q}^T_{2a})-q_{1a}^TC\stackrel{\leftrightarrow}{D}_\mu\gamma_5
q_{2b}(\bar{q}_{1a}\gamma^\mu \gamma_5C\bar{q}_{2b}^T-\bar{q}_{1b}
\gamma^\mu \gamma_5C\bar{q}^T_{2a}).\label{equ:current}
\end{split}
\end{equation}

%
%
\section{QCD sum rule}\label{sec:ope}
Consider the  two-point correlation function in the framework of
QCD sum rule
\begin{equation}
\Pi(q^{2})\equiv i \int d^4
xe^{iqx}\langle0|T\eta(x)\eta^\dag(0)|0\rangle, \label{equ:po}
\end{equation}
where $\eta$ is an interpolating current. At the hadron level, the
correlation function $\Pi(p^{2})$ is expressed via the dispersion
relation:
\begin{equation}
\Pi(p^2)=\int_{0}^{\infty}\frac{\rho(s)}{s-p^2-i\varepsilon}ds,\label{equ:pq}
\end{equation}
where
\begin{equation}
\begin{split}
\rho(s)\equiv&\sum_{n}\delta(s-m^{2}_{n})\langle0|\eta|n\rangle\langle
n|\eta^{\dag}|0\rangle\\
=&f^{2}_{X}\delta(s-m^{2}_{X})+\mbox{continuum} \; ,
\label{equ:rho}
\end{split}
\end{equation}
where $m_X$ is the mass of the resonance $X$ and $f_X$ is the
decay constant of the meson:
\begin{equation}
\langle0|\eta|X\rangle=f_X.
\end{equation}

The correlation function can also be calculated at the quark-gluon
level using the QCD operator product expansion (OPE) method. It is
convenient to evaluate the Wilson coefficient in the coordinate
space for the light quark systems and in the momentum space for
the heavy quark systems respectively. In our calculation we
consider the first order perturbative and various condensates
contributions. In order to calculate the gluonic condensate, it is
convenient to work in the fixed-point gauge. The massive quark
propagator $iS(x, y)$ in an external field in the fixed-point
gauge is listed in Appendix A. The quark lines attached with gluon
contain terms proportional to $y$, which we can ignore in the
current without derivatives. We keep these terms throughout the
evaluation and let $y$ go to zero only after finishing the
derivatives. The $\Pi (p^2)$ can be written as:
\begin{equation}
\Pi^{OPE}(p^2)=\int^{\infty}_{4(m_1+m_2)^2} ds
\frac{\rho^{OPE}(s)}{s-p^2-i \epsilon} ,
\end{equation}
where the $m_1$ and $m_2$ are the mass of the quark $q_1$ and
$q_2$ respectively. In order to suppress the higher state
contributions, we perform the Borel transformation to the
correlation function, which improves the convergence of the OPE
series. With the quark-hadron duality, we obtain:
\begin{equation}
\Pi(M_B^2)=f_X^2e^{-m_X^2/ M_B^2}=\int_{4(m_1+m_2)^2}^{\infty} ds
e^{-s/M_B^2}\rho^{OPE}(s),
\end{equation}
where $s_0$ is the threshold parameter and $M_B$ is the Borel
parameter. We can extract the meson mass $m_X$:
\begin{equation}
m_X^2=-\frac{\frac{\partial}{\partial (1/M_B^2)}
\Pi(M_B^2)}{\Pi(M_B^2)}=\frac{\int_{4(m_1+m_2)^2}^{s_0} ds
e^{-s/M_B^2} s \rho(s)}{\int_{4(m_1+m_2)^2}^{s_0} ds e^{-s/M_B^2}
\rho(s)}
\end{equation}

\begin{figure}[hbtp]
\begin{center}
\scalebox{0.50}{\includegraphics{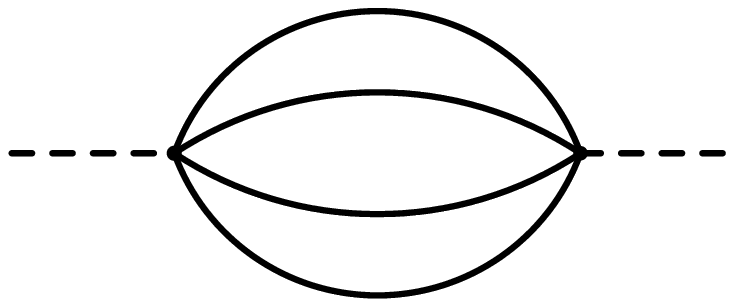}}\quad
\scalebox{0.50}{\includegraphics{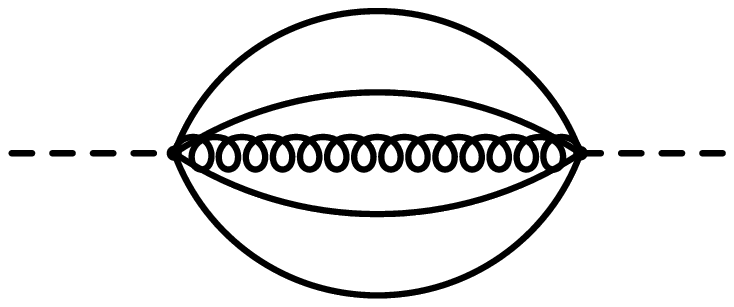}}\quad
\scalebox{0.50}{\includegraphics{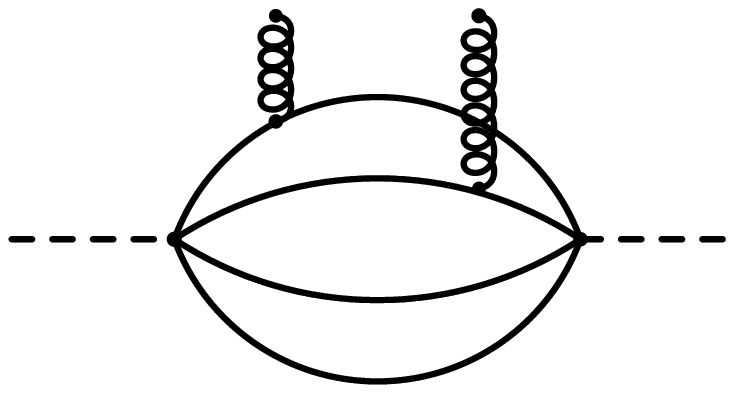}}\\
\scalebox{0.50}{\includegraphics{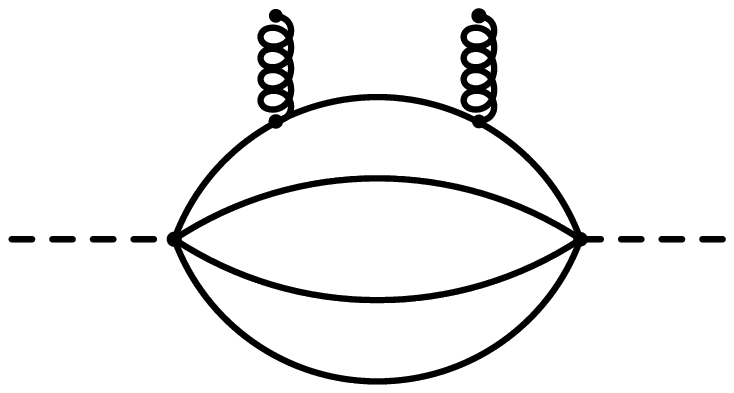}}\quad
\scalebox{0.50}{\includegraphics{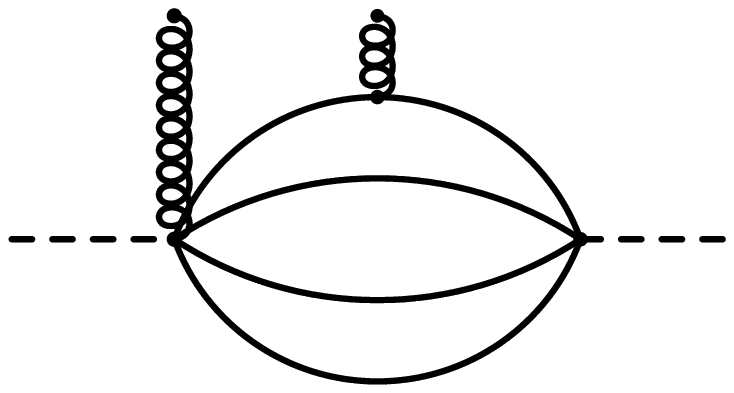}}\quad
\scalebox{0.50}{\includegraphics{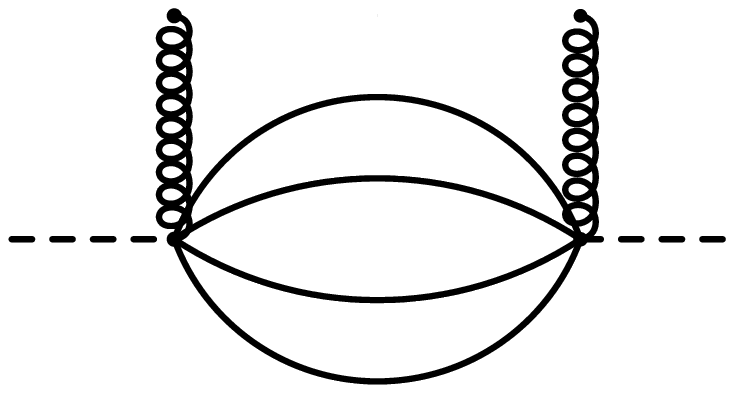}}\\
\scalebox{0.50}{\includegraphics{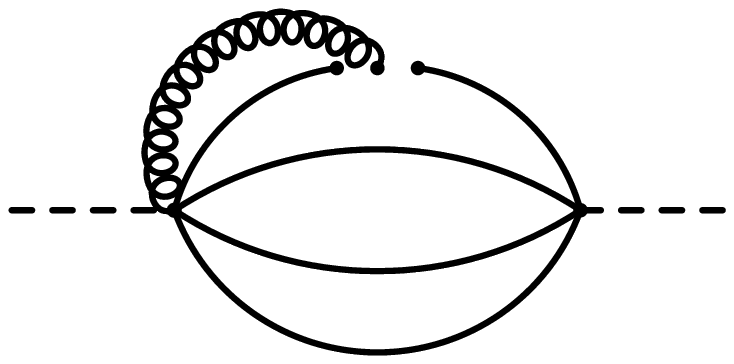}}\quad
\scalebox{0.50}{\includegraphics{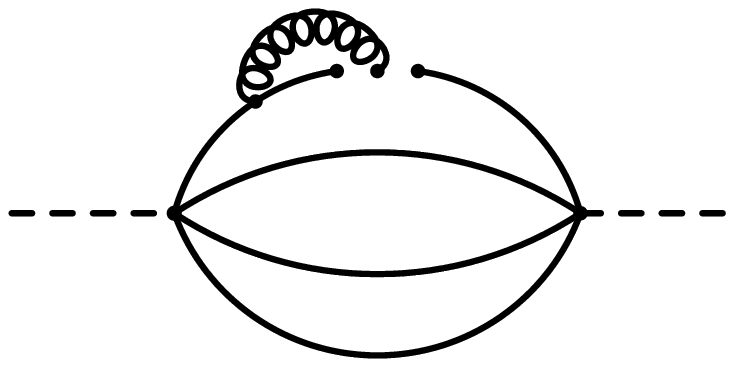}}\quad
\scalebox{0.50}{\includegraphics{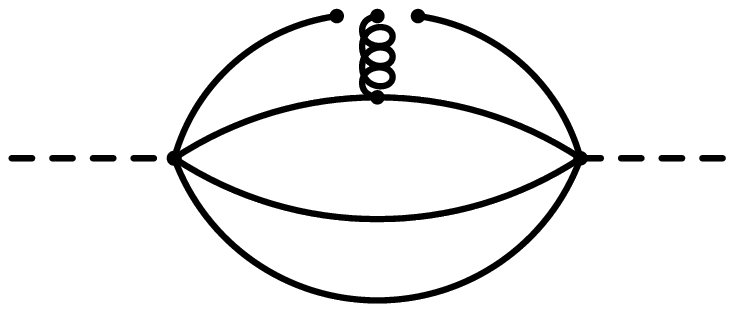}} \caption{Some typical Feynman diagrams of the correlation fuctions.} \label{feyn1}
\end{center}
\end{figure}

For all the tetraquark currents in~Eq.(\ref{equ:current}), we
collect the spectral density $\rho^{OPE}(s)$ in the Appendix. The
quark condensate $\langle\bar{q}q\rangle$ vanishes due to the
special Lorenz structures of the currents. For the $q=u,d$, we do
the calculation in the chiral limit $m_q=0$. Since the
contribution of the three gluon condensate $\langle
g^2fGGG\rangle$ is very small, we consider only the power
corrections from the following condensates $\langle g^2
GG\rangle$, $\langle\bar{q}g\sigma \cdot G q\rangle$,
$\langle\bar{q}q\rangle^2$ and $\langle\bar{q}g\sigma \cdot G
q\rangle \langle\bar{q}q\rangle$. We list several typical Feynman
diagrams in the Fig. \ref{feyn1}.

%
%

%

\section{Numerical Analysis}\label{sec:num}
%
In the QCD sum rule analysis,we use the following values of the
quark masses, coupling constant and various
condensates~\cite{Amsler:2008zzb,Shifman:1978bx,eidemuller,Narison}:
\begin{equation}
\begin{split}
&m_c(m_c)=(1.23\pm0.09)~\mbox{GeV},\\
&m_b(m_b)=(4.20\pm0.07)~\mbox{GeV},\\
&\langle \bar{q}q\rangle=-(0.23\pm0.03)^3~\mbox{GeV}^3,\\
&\langle \bar{q}g \sigma\cdot G q\rangle=-M_0^2\langle \bar{q}q\rangle,\\
&M_0^2=(0.8\pm0.2)~\mbox{GeV},\\
&\langle \bar{s}s\rangle/ \langle\bar{q}q\rangle=0.8\pm0.2,\\
&\langle g^2 GG\rangle=(0.88\pm0.13)~\mbox{GeV}^4,\\
&\alpha_s(1.7\mbox{GeV})=0.328\pm0.03\pm0.025.
\end{split}
\end{equation}

The Borel mass $M_B$ and the threshold value $s_0$ are two pivotal
parameters. The working region of the Borel mass is determined by
the convergence of the OPE and the pole contribution. The
requirement of the convergence of the OPE determines the lower
bound $M_{Bmin}$ of the Borel mass, and the pole contribution
determines the upper bound $M_{Bmax}$.

In this work, there is no contribution from the quark condensate
$\langle \bar{q} q\rangle $. The correction from the condensate
$\langle \bar{q}g\sigma \cdot G q\rangle \langle \bar{q} q\rangle$
is the most important numerically. Its contribution is bigger
than that from the gluon condensate $\langle g^2 GG\rangle$, mixed
condensate $\langle \bar{q}g\sigma \cdot G q\rangle$ and the four
quark condensate $\langle \bar{q}q\rangle^2$. The mixed condensate
is also very important numerically for the currents $\eta_{1,3}$.
We require that the condensate $\langle \bar{q}g\sigma \cdot G
q\rangle \langle \bar{q} q\rangle$ be less than one ninth of the
perturbative term to ensure the convergence of OPE, which leads to
the lower bound of the Borel parameter working window.

The pole contribution (PC) is defined as
\begin{equation}
\mbox{PC}=\frac{\int_{4(m_1+m_2)^2}^{s_0} ds
e^{-s/M_B^2}\rho(s)}{\int_{4(m_1+m_2)^2}^{\infty} ds
e^{-s/M_B^2}\rho(s)}
\end{equation}
which depends on both the Borel mass $M_B$ and the threshold value
$s_0$. $s_0$ is chosen around the region where the variation of
$m_X$ with $M_B$ is minimum. Requiring the PC be larger than $30\%
\sim 50\%$, we get the upper bound $M_{Bmax}$ of the Borel mass
$M_B$. We list the working region of the Borel parameter for the
four currents with different quark composition in Table
\ref{tab1}. For the $\eta^c_{1-4}$, we get the upper bound of
Borel parameter $M_B$ by requiring the PC be larger than $30\%$.
For $\eta^b_{1-4}$ we require the PC be larger than $40\%$. The
masses are extracted using the threshold values $s_0$ and Borel
parameters $M_B$ listed in Table \ref{tab1}. The last column is
the pole contribution with the corresponding $s_0$ and $M_B$.
\begin{table}
\begin{tabular}{c|c|c|c|c|c|c}
  \hline
  \hline
    & Current & $s_0$($\mbox{GeV}^2$) & [$M_{Bmin},M_{Bmax}$]($\mbox{GeV}$) & $M_B$(GeV) & $m_X (\mbox{GeV})$ & PC(\%) \\
  \hline

  $q_1,q_2=u,d$ & $\eta^q_{1-4}$ & - & - & - & - & - \\
  \hline
    & $\eta^c_1$ & 27 & $1.8\sim 2.1$ & 2.0 & $4.76\pm0.08$  & 37.4  \\
  $q_1=u,d$ & $\eta^c_2$ & 28 & $1.8\sim 2.1$ & 2.0 & $4.85\pm0.09$ & 39.9  \\
  $q_2=c$  & $\eta^c_3$ & 29 & $1.8\sim 2.1$ & 2.0 & $4.96\pm0.13$ & 42.4 \\
    & $\eta^c_4$ & 28 & $1.8\sim 2.1$ & 2.0 & $4.83\pm0.07$ & 40.9 \\
  \hline
    & $\eta^b_1$ & 140 & $2.9\sim 3.3$ & 3.1 & $11.24\pm0.17$ & 52.2 \\
  $q_1=u,d$ & $\eta^b_2$ & 142 & $2.9\sim 3.3$ & 3.1 & $11.27\pm0.14$ & 55.6  \\
  $q_2=b$  & $\eta^b_3$ & 142 & $2.9\sim 3.3$ & 3.1 & $11.30\pm0.17$  & 55.0 \\
    & $\eta^b_4$ & 142 & $2.9\sim 3.3$ & 3.1 & $11.27\pm0.09$ & 55.5  \\
  \hline
  \hline
\end{tabular}
\caption{The threshold values, Borel window, Borel parameter for
the different tetraquark currents. \label{tab1}}
\end{table}

\begin{figure}[hbtp]
\begin{center}
\scalebox{0.75}{\includegraphics{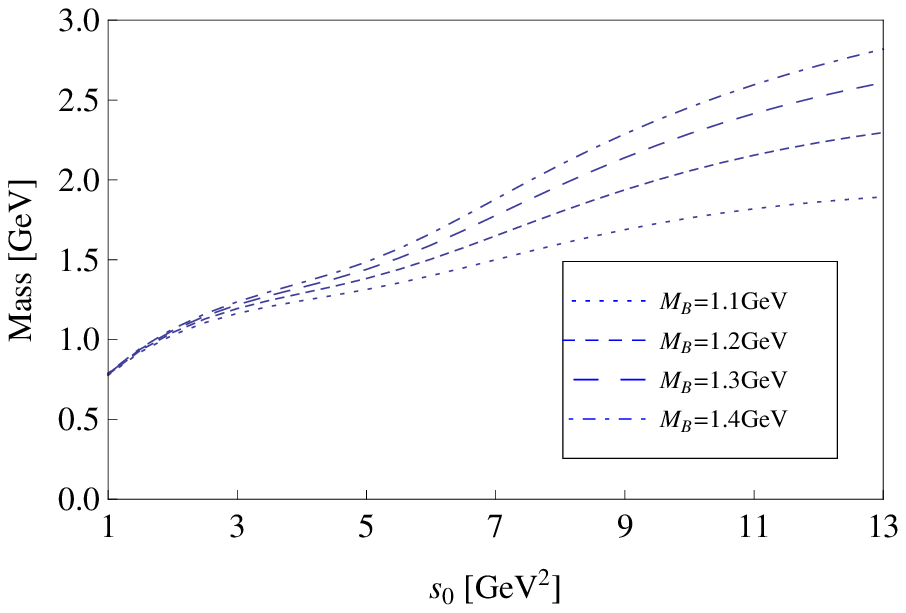}}\quad
\scalebox{0.75}{\includegraphics{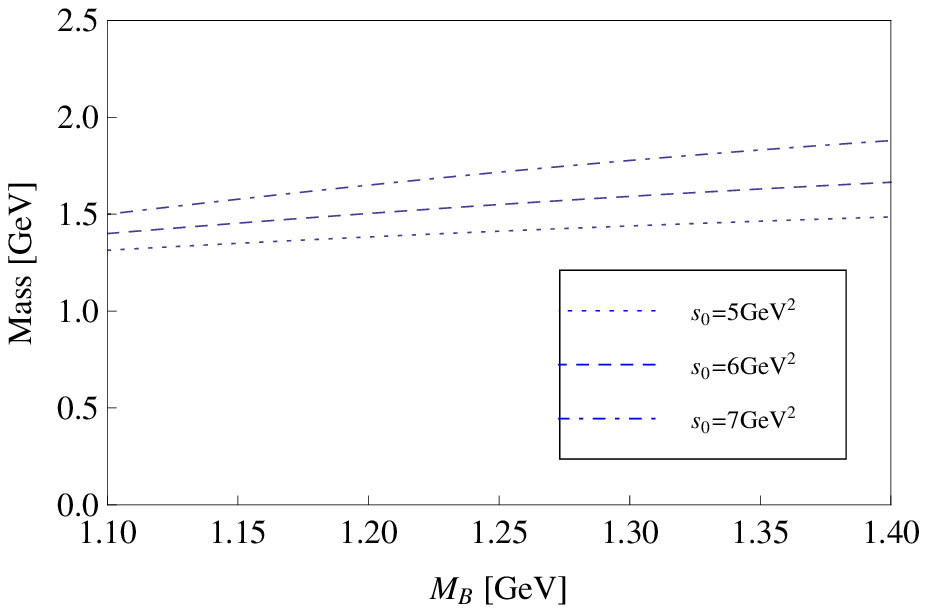}} \caption{The
variation of $m_X$ with $M_B$ (left) and $s_0$ (right) for the
current $\eta^q_1$.}\label{figq1}
\scalebox{0.75}{\includegraphics{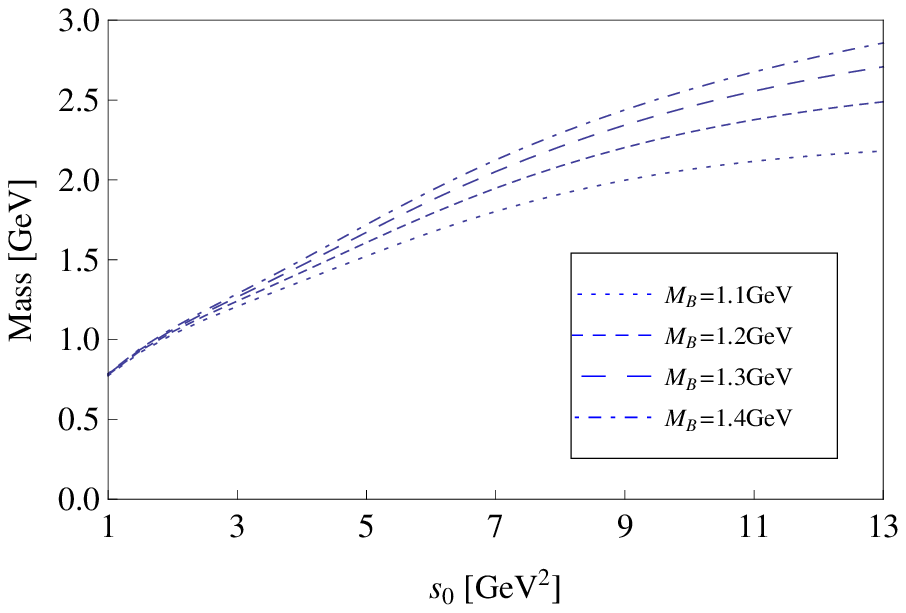}}\quad
\scalebox{0.75}{\includegraphics{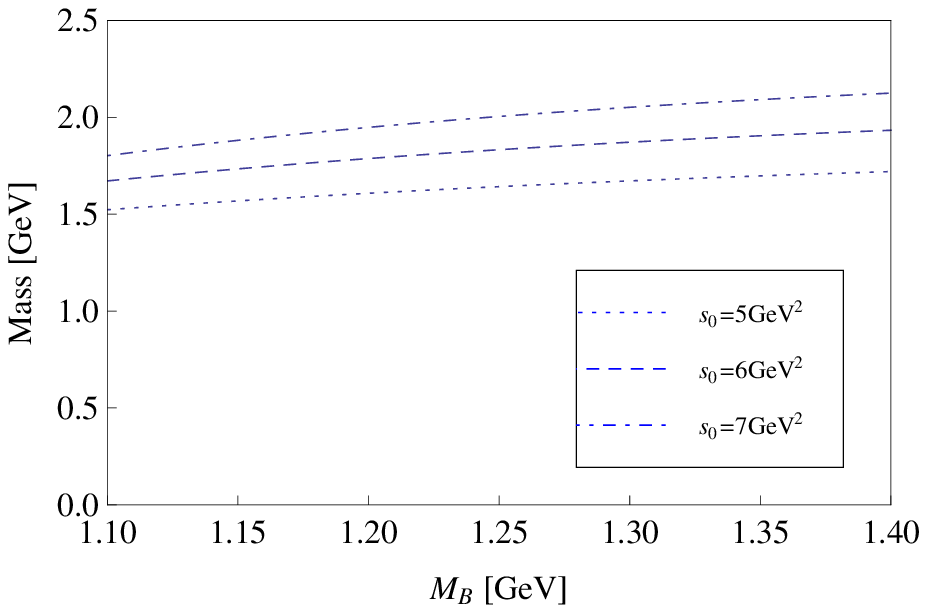}} \caption{The
variation of $m_X$ with $M_B$ (left) and $s_0$ (right) for the
current $\eta^q_2$.}\label{figq2}
\end{center}
\end{figure}

For the light tetraquark systems, there does not exist a working
region for the sum rules. Even in the extreme case that the pole
contribution is $\sim 30\%$ and the contribution of the condensate
$\langle \bar{q}g\sigma\cdot Gq\rangle\langle\bar{q}q\rangle$ is
around the leading order contribution, the lower bound $M_{Bmin}$
is still much larger than the upper bound $M_{Bmax}$. In other
words, there is no working region for light quark systems. As
shown in~Fig.(\ref{figq1},\ref{figq2}), the extracted mass grows
monotonically with $s_0$ which implies the $0^{+-}$ state does not
exist below 2 GeV. We note that the light $J^{PC}=0^{--}$ state
does not exist either~\cite{Jiao:2009}. The $0^{+-}$ and $0^{-+}$
channels are in strong contrast with the $0^{++}$ case, where
there exist stable tetraquark QCD sum rules and the extracted
scalar meson masses agree with the experimental scalar spectrum
nicely \cite{CHX}.

\begin{figure}[hbtp]
\begin{center}
\scalebox{0.75}{\includegraphics{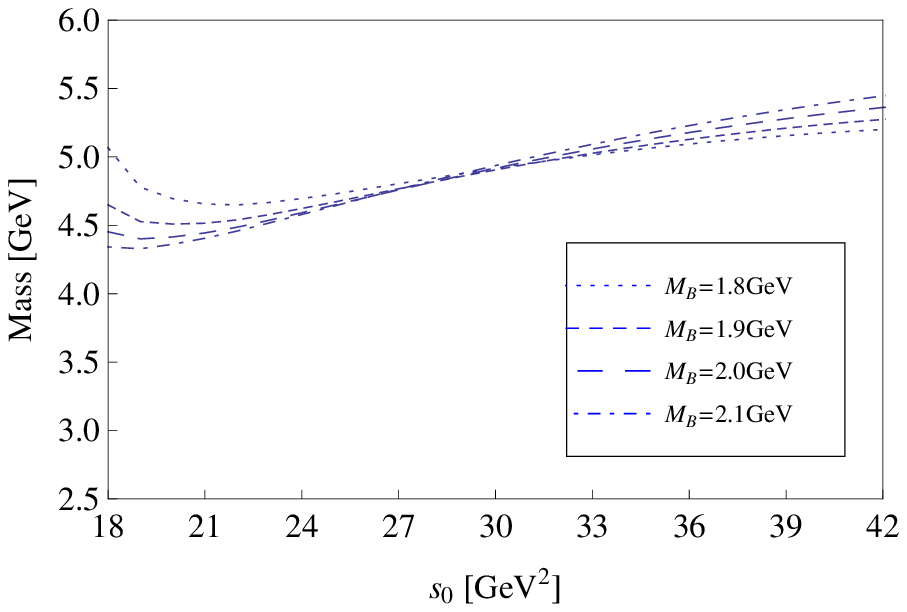}}\quad
\scalebox{0.75}{\includegraphics{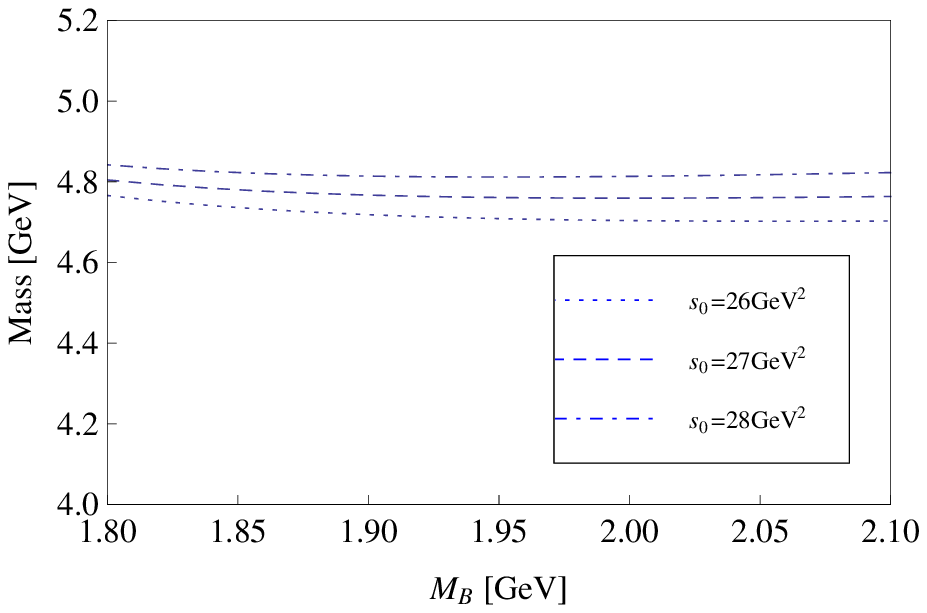}}\\
\caption{The variation of $m_X$ with $M_B$ (left) and $s_0$
(right) for the current $\eta^c_1$.}\label{XX1}
\scalebox{0.75}{\includegraphics{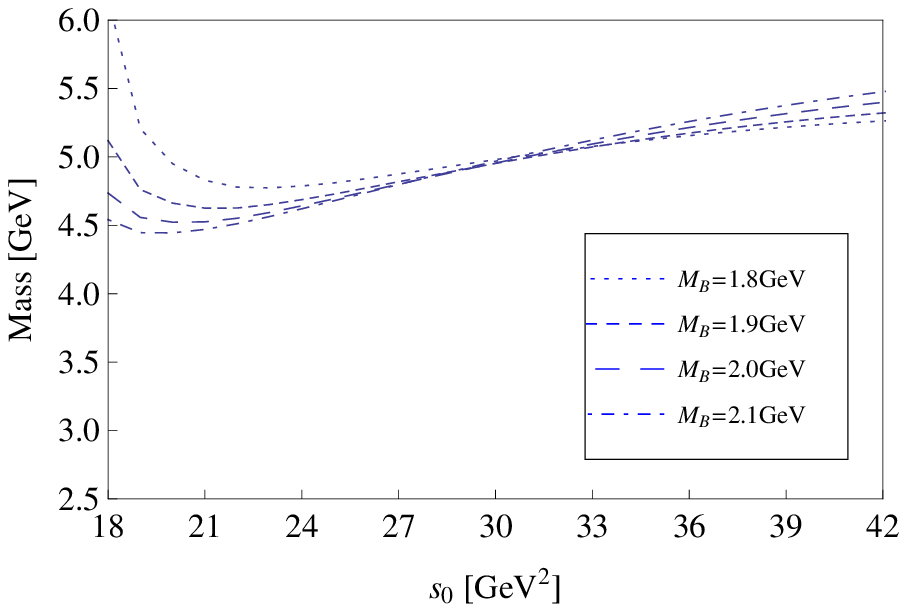}}\quad
\scalebox{0.75}{\includegraphics{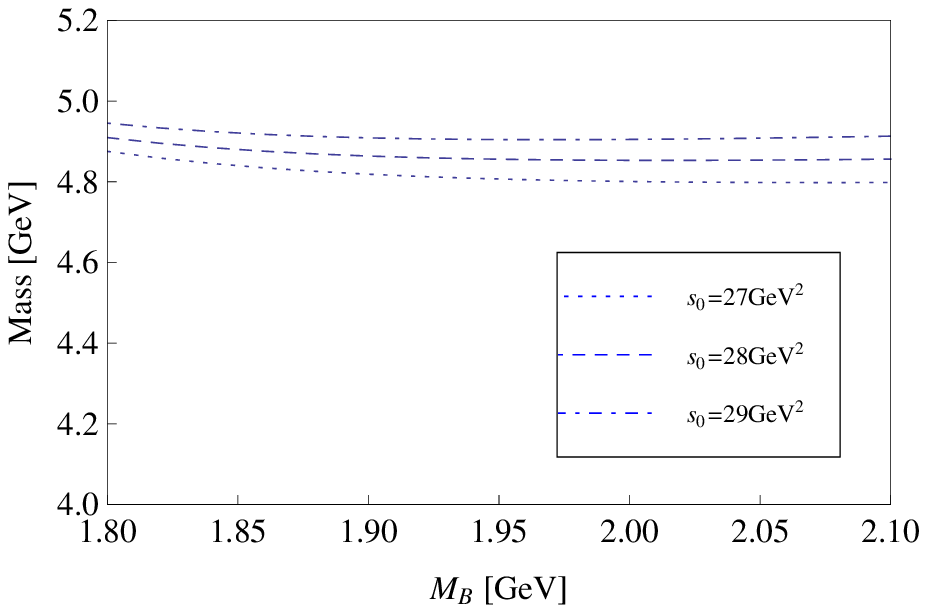}}\\
\caption{The variation of $m_X$ with $M_B$ (left) and $s_0$
(right) for the current $\eta^c_2$.}
\scalebox{0.75}{\includegraphics{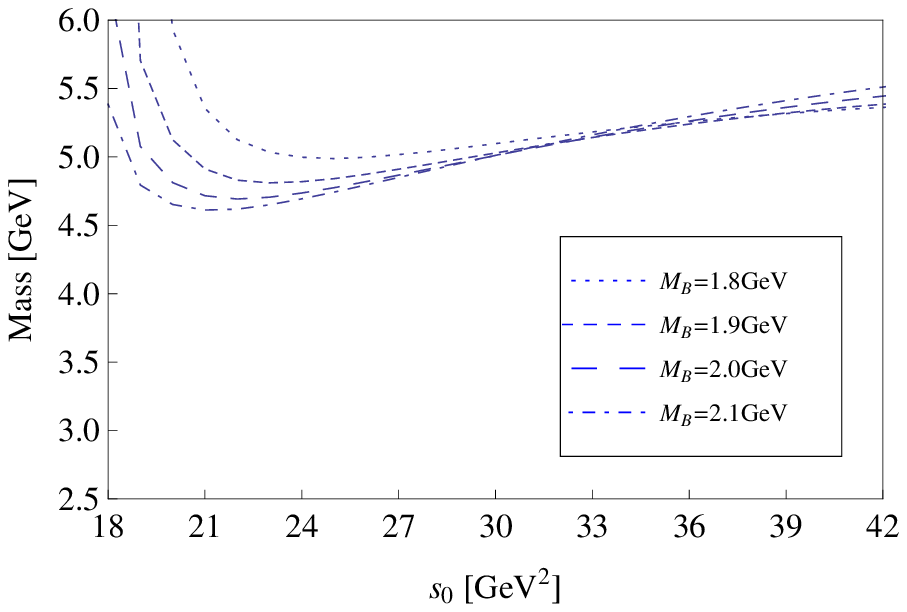}}\quad
\scalebox{0.75}{\includegraphics{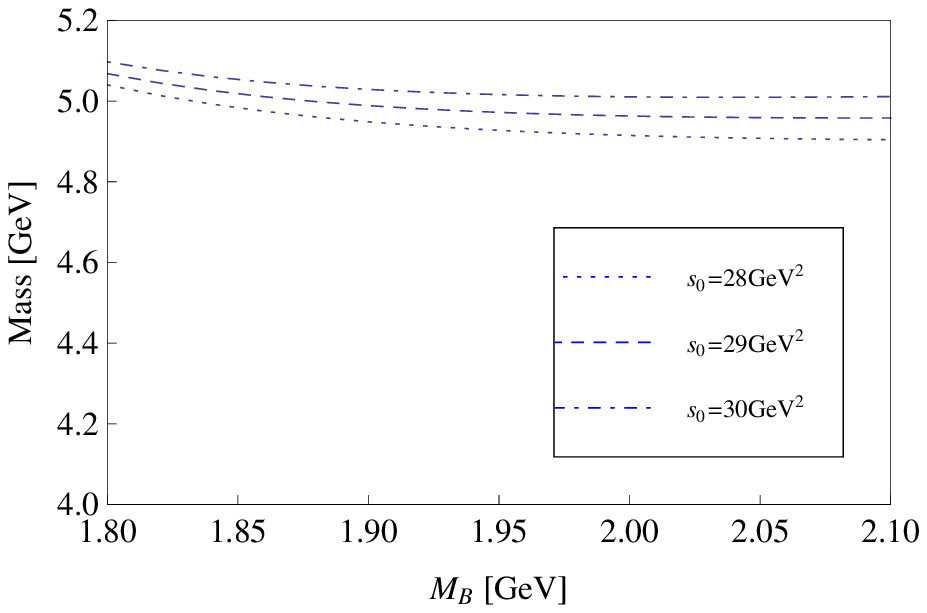}}\\
\caption{The variation of $m_X$ with $M_B$ (left) and $s_0$
(right) for the current $\eta^c_3$.}
\scalebox{0.75}{\includegraphics{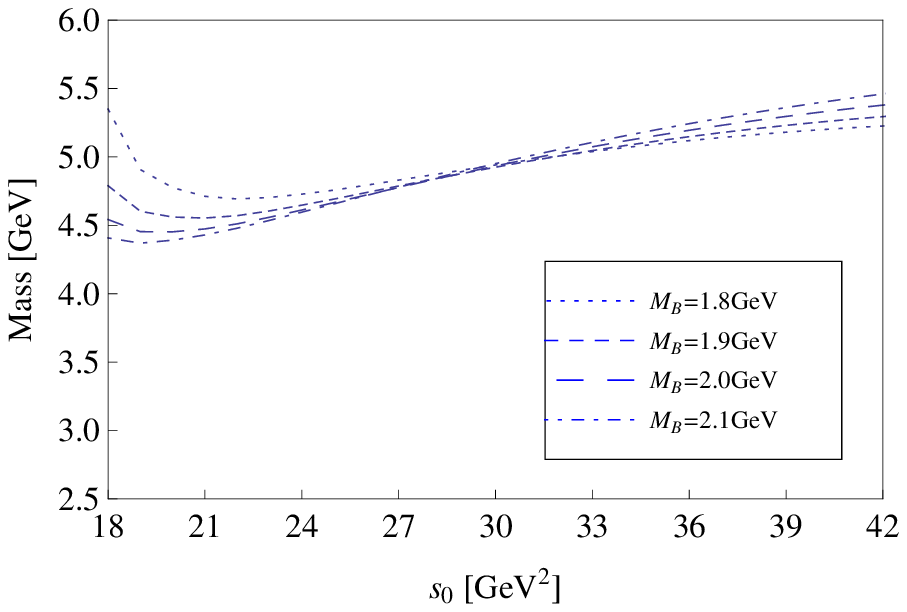}}\quad
\scalebox{0.75}{\includegraphics{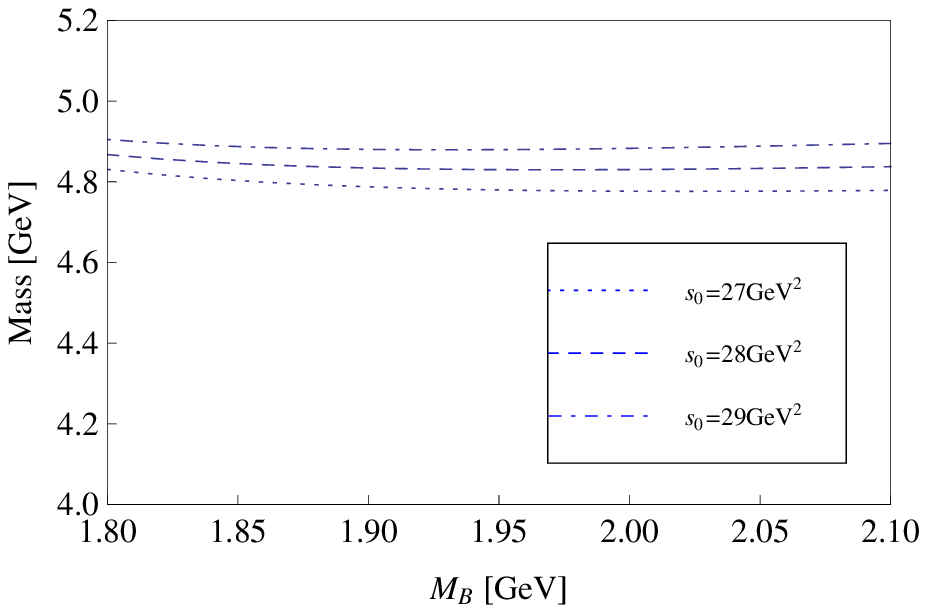}}\\
\caption{The variation of $m_X$ with $M_B$ (left) and $s_0$
(right) for the current $\eta^c_4$.}
\end{center}
\end{figure}

\begin{figure}[hbtp]
\begin{center}
\scalebox{0.75}{\includegraphics{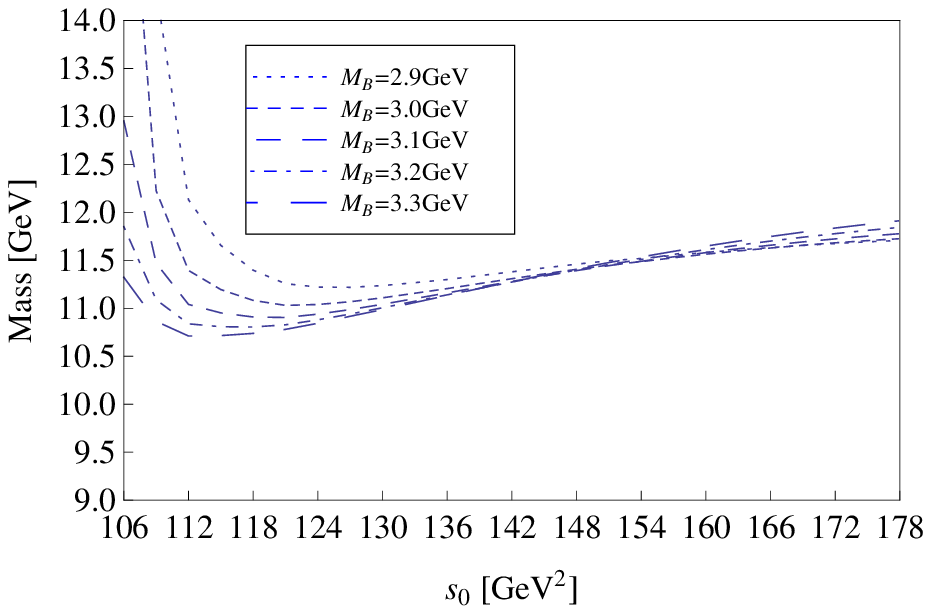}}\quad
\scalebox{0.75}{\includegraphics{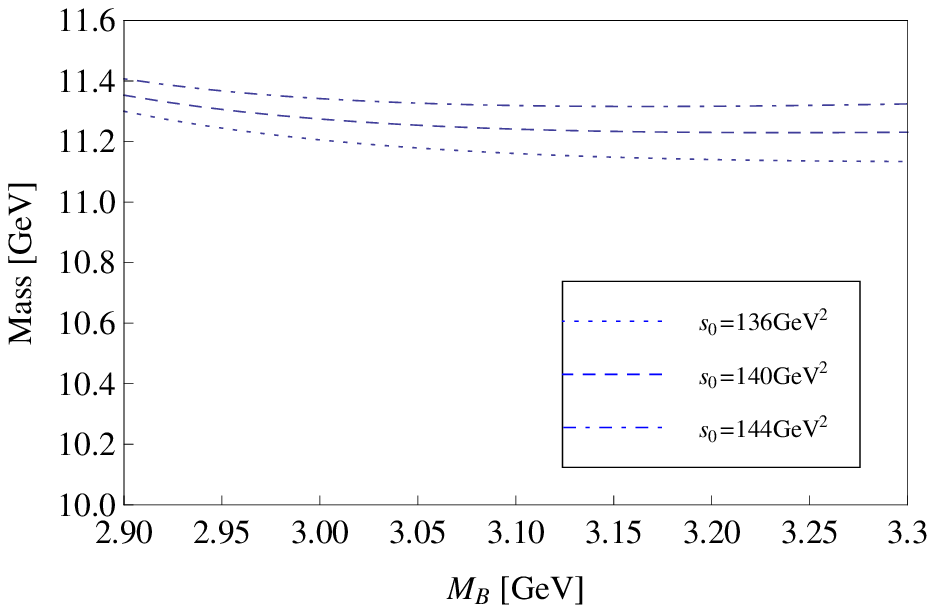}}\\
\caption{The variation of $m_X$ with $M_B$ (left) and $s_0$
(right) for the current $\eta^b_1$.}
\scalebox{0.75}{\includegraphics{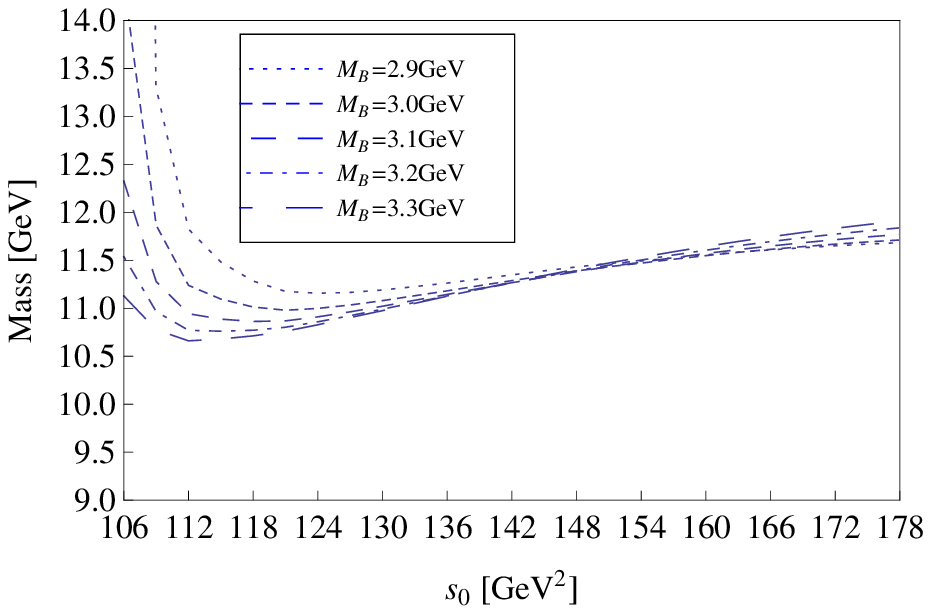}}\quad
\scalebox{0.75}{\includegraphics{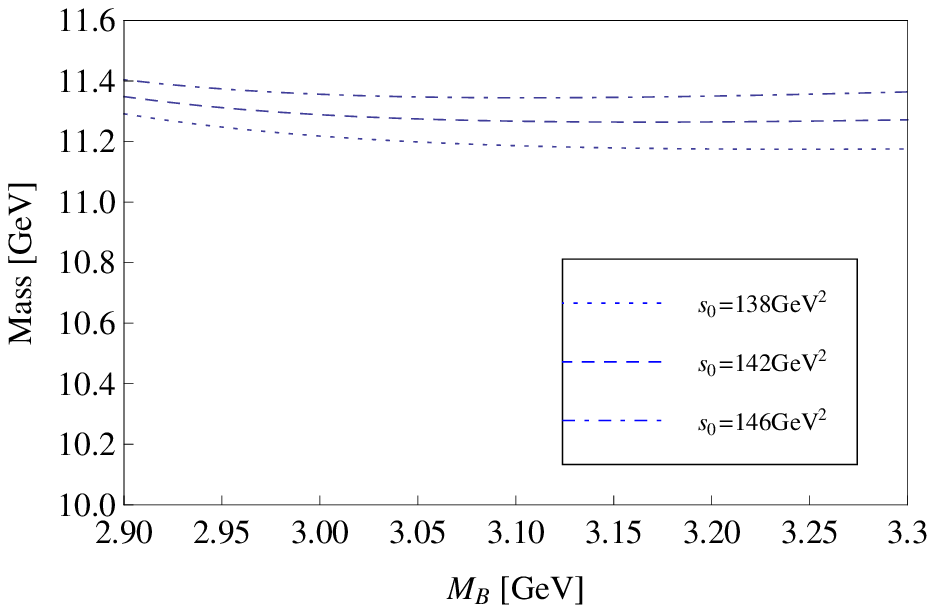}}\\
\caption{The variation of $m_X$ with $M_B$ (left) and $s_0$
(right) for the current $\eta^b_2$.}
\scalebox{0.75}{\includegraphics{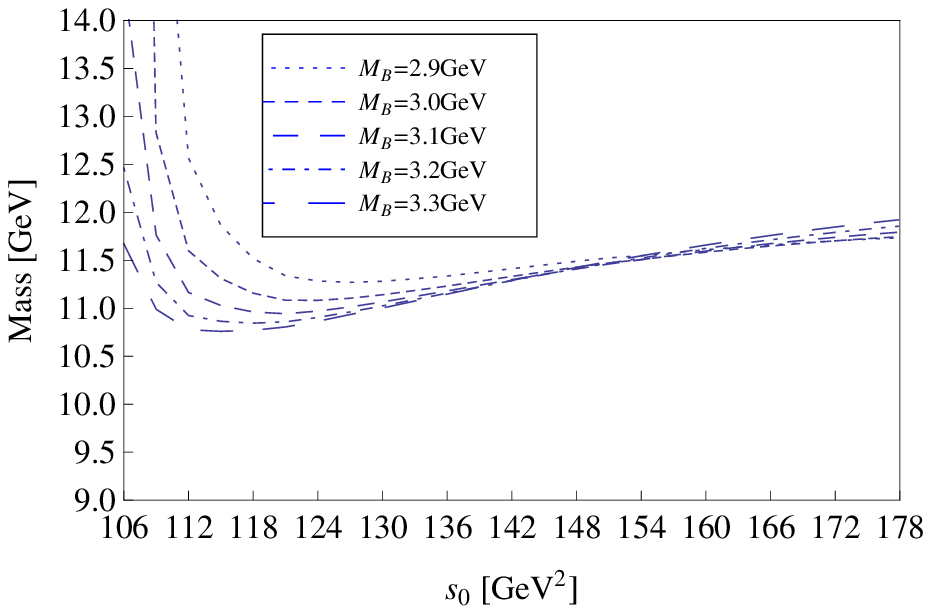}}\quad
\scalebox{0.75}{\includegraphics{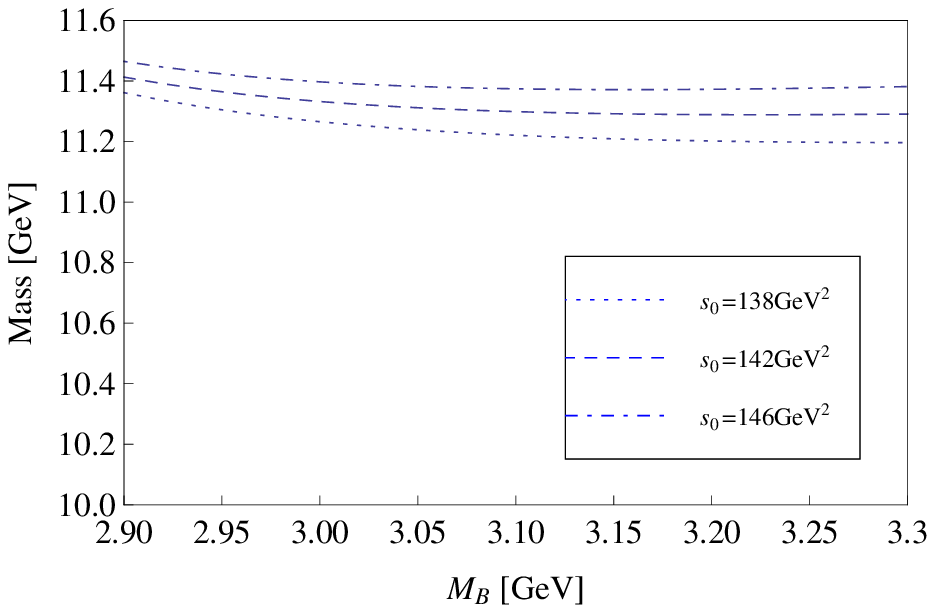}}\\
\caption{The variation of $m_X$ with $M_B$ (left) and $s_0$
(right) for the current $\eta^b_3$.}
\scalebox{0.75}{\includegraphics{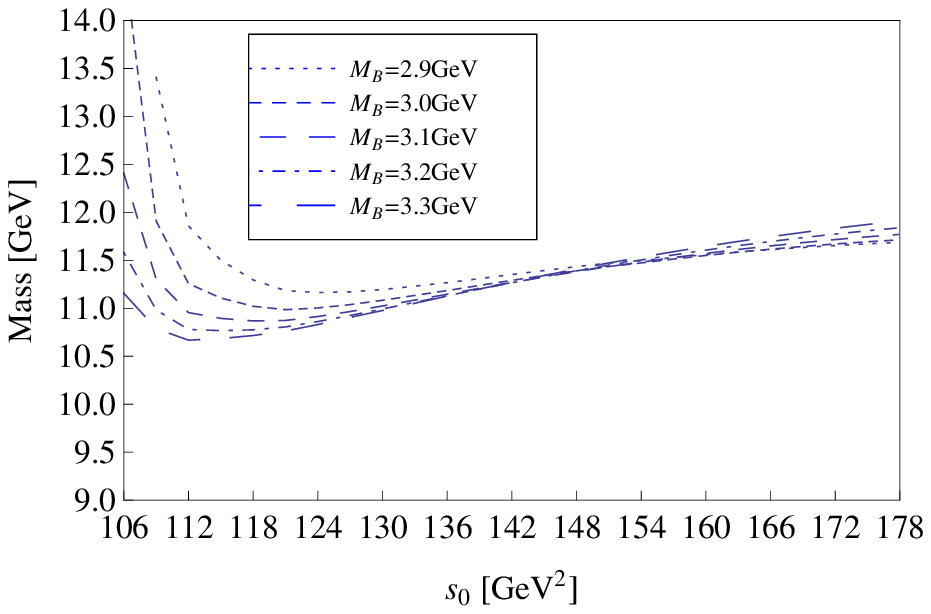}}\quad
\scalebox{0.75}{\includegraphics{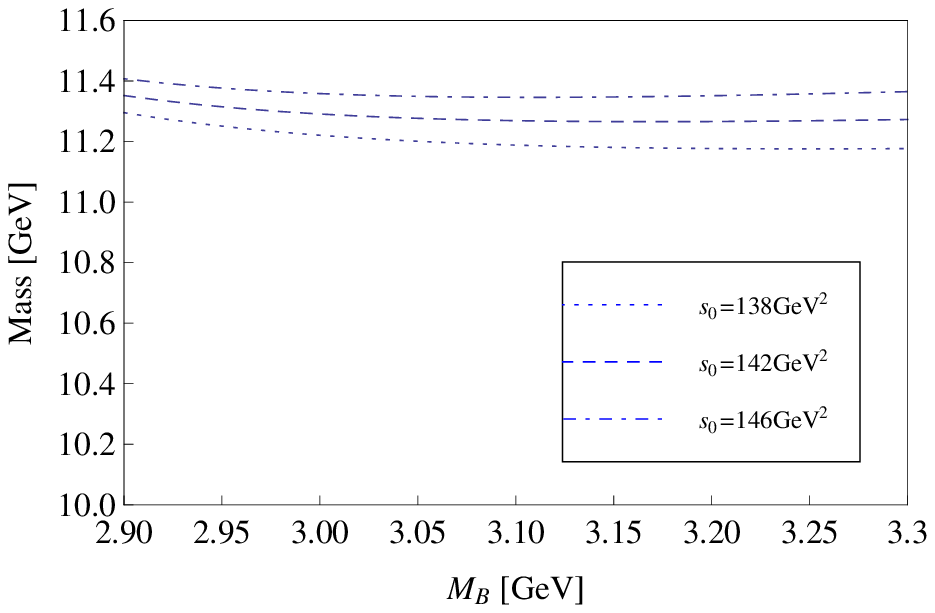}}\\
\caption{The variation of $m_X$ with $M_B$ (left) and $s_0$
(right) for the current $\eta^b_4$.}\label{xx2}
\end{center}
\end{figure}

For the heavy systems, the variation of $m_X$ with $s_0$ and $M_B$
is presented in Figs. (\ref{XX1})-(\ref{xx2}). All the sum rules
are very stable with reasonable variations of $s_0$ and $M_B$. The
presence of the two heavy quarks reduces the kinetic energy of the
tetraquark system, hence helps to stabilize the sum rules.
Numerically, the masses of the $0^{+-}$ states are slightly larger
than those of the $0^{--}$ states ~\cite{Wei:2010}.

\section{Summary}\label{sec:summary}

The exotic state with $J^{PC}=0^{+-}$ cannot be composed of a pair
of quark and anti-quark. In order to explore these exotic states,
we have constructed four tetraquark interpolating operators. Then
we make the operator product expansion and extract the spectral
density. Because of the special Lorentz structures of the
currents, the quark condensate $\langle\bar{q}q\rangle$ vanishes.

For the light tetraquark systems, there does not exist a working
region of the Borel parameter and threshold for all the derived
sum rules. It seems that none of these independent interpolating
currents supports a resonant signal below 2 GeV, which is
consistent with the current experimental data
\cite{Amsler:2008zzb}. In contrast, there exist very stable QCD
sum rules constructed from the tetraquark interpolating operators
in the scalar channel. The extracted scalar spectrum agrees with
the experimental data nicely \cite{CHX}.

For the heavy quark systems, the $0^{+-}$ tetraquark sum rules are
quite stable. The presence of two heavy quarks may render the
kinetic energy of the tetraquark system, which is helpful in the
formation of bound states. The extracted masses from the four
interpolating currents $\eta^c_{1-4}$ are around $4.76\sim 4.96~
\mbox{GeV}$ for the charmonium-like states. For the
bottomnium-like $0^{+-}$ states, their masses are about $11.2\sim
11.3~\mbox{GeV}$. It's very interesting to note that the mass of
the $0^{+-}$ charmonium-like state extracted from the tetraquark
sum rules is numerically quite close to the mass of the $0^{+-}$
hybrid charmonium extracted on the lattice \cite{lat1,lat2}.

Because of the special ``exotic" quantum number, the $0^{+-}$
charmonium-like state does not decay into a pair of particle (H)
and anti-particle ($\bar H$). There are two types of $0^{+-}$
states with different isospin and G-parity: $I^G=0^-$ and
$I^G=1^+$. Only a few S-wave decay modes are allowed. Some
 possible two-body decay modes are listed in Table~\ref{decay}.
  Replacing the D meson by B meson, one gets the
decay patterns of the bottomonium-like states so long as the
kinematics allows. The $0^{+-}$ state may be searched for
experimentally at facilities such as Super-B factories, PANDE, LHC
and RHIC in the future, especially at RHIC and LHC where plenty of
charm, anti-charm and light quarks are produced.

\begin{table}
\begin{tabular}{c|c|c}
  \hline
  \hline
  $I^G$ & S-wave & P-wave \\
  \hline
    &  & $D^0(1865)\bar{D}_1(2420)^0+c.c.$,$D^*(2007)^0\bar{D}_0^*(2400)^0+c.c.$,  \\
  $0^-$ & $\chi_{c1}(1P)h_1(1170)$\dots & $\eta_c(1S)h_1(1170),J/\psi(1S)f_0(600)$, \\
    &  & $J/\psi f_0(980),J/\psi f_1(1285),\chi_{c0}(1P)\omega(782),\chi_{c1}(1P)\omega(782)$,\\
    &   & $\psi(2S)f_0(600),\psi(3770)f_0(600)\dots$ \\
  \hline
    & $J/\psi(1S)\pi_1(1400)$,  & $D^0(1865)\bar D_1(2420)^0+c.c.,D^*(2007)^0\bar{D}^*_0(2400)^0+c.c.,$ \\
  $1^+$  & $J/\psi(1S)\pi_1(1600)$,  & $D^*(2007)^0\bar D_1(2420)^0+c.c.$,\\
    & $\chi_{c1}(1P)b_1(1235)\dots$   & $\eta_c(1S)b_1(1235),J/\psi(1S)a_0(980),J/\psi(1S)a_1(1260)$, \\
    &  &$\chi_{c0}(1P)\rho(770),\chi_{c1}(1P)\rho(770)\dots$ \\
  \hline
  \hline
\end{tabular}
\caption{The possible decay modes of the $0^{+-}$ charmonium-like state.\label{decay}}
\end{table}

\section*{Acknowledgments}

This project was supported by the National Natural Science
Foundation of China under Grants 11075004, 11021092 and Ministry
of Science and Technology of China (2009CB825200).


\appendix

\section{The Momentum Space Propagator}\label{app}
The fixed-point gauge is defined as:
\begin{equation}
(x-x_0)^\mu A^a_\mu(x)=0
\end{equation}
where $x_0$ is an arbitrary point in space which can be chosen to
be the origin. Then the potential $A_\mu^a$ can be expressed in
terms of the field strength tensor
$G_{\mu\nu}$($G_{\mu\nu}=\frac{\lambda^a}{2}G^a_{\mu\nu}$)~\cite{Reinders:1984sr,Wei:2011}:
\begin{equation}
\begin{split}
A_\mu(x)=&\int^1_0 tdt G_{\nu\mu}(tx)x^\nu \\
&\frac{1}{2}x^\nu G_{\nu\mu}(0)+\frac{1}{3}x^\alpha x^\nu D_\alpha
G_{\nu\mu}(0)+\frac{1}{8}x^\alpha x^\beta x^\nu D_\alpha D_\beta
G_{\nu\mu}(0)+\dots ,
\end{split}
\end{equation}
Denote the massive quark propagator between the position $x$ and
$y$ in the coordinate space as $iS(x,y)$. The massive quark
propagator in the momentum space is ~\cite{Wei:2011}:
\begin{equation}
iS(p)=iS_0(p)+iS_g(p)+iS_{gg}(p)+\dots,
\end{equation}
where $iS_0(p)$ is the free quark propagator:
\begin{equation}
iS_0(p)=\frac{i}{\hat{p}-m},
\end{equation}
where $\hat{p}=\gamma^\mu p_\mu$, $iS_g(p)$ is the quark
propagator with one gluon leg attached:
\begin{equation}
iS_g(p)=\frac{i}{4}\frac{\lambda^n}{2}g_sG^n_{\mu\nu}\frac{\sigma^{\mu\nu}(\hat{p}+m)+(\hat{p}+m)\sigma^{\mu\nu}}{(p^2-m^2)^2}
+\frac{i}{2}\frac{\lambda^n}{2}g_sG^n_{\mu\nu} \Big[ \frac{2y^\mu
p^\nu (\hat{p}+m)}{(p^2-m^2)^2}-\frac{y^\mu
\gamma^\nu}{p^2-m^2}\Big]
\end{equation}
$iS_{gg}(p)$ is the quark propagator with two gluon legs attached:
\begin{equation}
\begin{split}
iS_{gg}(p)=&-\frac{i}{4}\frac{\lambda^a}{2}\frac{\lambda^b}{2}g_s^2 G^a_{\mu\rho}G^b_{\nu\sigma}\frac{\hat{p}+m}{(p^2-m^2)^5}(f^{\mu\rho\nu\sigma}+f^{\mu\nu\rho\sigma}+f^{\mu\nu\sigma\rho})\\
&-\frac{1}{4}\frac{\lambda^a}{2}\frac{\lambda^b}{2}g_s^2
G^a_{\mu\rho}G^b_{\nu\sigma}\frac{\hat{p}+m}{(p^2-m^2)^4}\Big[
y^\sigma (f^{\mu\rho\nu}+f^{\mu\nu\rho})+y^\rho
f^{\mu\nu\sigma}-iy^{\rho}y^\sigma (p^2-m^2)\Big]
\end{split}
\end{equation}
where $f^{\mu\nu\dots
\alpha\beta}=\gamma^\mu(\hat{p}+m)\gamma^\nu(\hat{p}+m)\dots
\gamma^\alpha(\hat{p}+m)\gamma^\beta(\hat{p}+m)$.

\section{The Spectral Densities}\label{sec:rhos}
%
In this appendix, we list the spectral densities of the tetraquark
interpolating currents. For the light quark systems
($q_1,q_2=u,d$), the spectral densities are:
\begin{equation}
\begin{split}
&\rho_1(s)=\frac{s^5}{51200 \pi ^6}(1+\frac{17}{108}\frac{\alpha}{\pi})-\frac{\langle g^2  GG\rangle s^3}{18432 \pi ^6}-\frac{\langle \bar{q}q\rangle^2 s^2}{6 \pi ^2}-\frac{799 \langle \bar{q}g_s\sigma \cdot Gq\rangle \langle \bar{q}q\rangle s}{768 \pi ^2}\\
&\rho_2(s)=\frac{s^5}{102400 \pi ^6}(1+\frac{7}{54}\frac{\alpha}{\pi})+\frac{ \langle  g^2GG\rangle s^3}{18432 \pi ^6}-\frac{\langle \bar{q}q\rangle^2 s^2}{12 \pi ^2}-\frac{245 \langle \bar{q}g_s\sigma \cdot Gq\rangle \langle \bar{q}q\rangle s}{768 \pi ^2}\\
&\rho_3(s)=\frac{s^5}{51200 \pi ^6}(1+\frac{17}{108}\frac{\alpha}{\pi})-\frac{\langle  g^2GG\rangle s^3}{18432 \pi ^6}-\frac{\langle \bar{q}q\rangle^2 s^2}{6 \pi ^2}-\frac{7 \langle \bar{q}g_s\sigma \cdot Gq\rangle \langle \bar{q}q\rangle s}{12 \pi ^2}\\
&\rho_4(s)=\frac{s^5}{102400 \pi
^6}(1+\frac{7}{54}\frac{\alpha}{\pi})+\frac{ \langle
g^2GG\rangle s^3}{18432 \pi ^6}-\frac{\langle \bar{q}q\rangle^2
s^2}{12 \pi ^2}-\frac{7 \langle \bar{q}g_s\sigma \cdot Gq\rangle
\langle \bar{q}q\rangle s}{24 \pi ^2}
\end{split}
\end{equation}

For the heavy systems ($q_1=u,d$,~$q_2=c,d$), the spectral
densities are:
\begin{equation}
\rho(s)=\rho^{pert}(s)+\rho^{\langle GG\rangle}(s)+\rho^{\langle
\bar{q}Gq\rangle}(s)+\rho^{\langle
\bar{q}q\rangle^2}(s)+\rho^{\langle \bar{q}Gq\rangle \langle
\bar{q}q\rangle}(s)
\end{equation}

For the condensate $\langle \bar qq\rangle \langle \bar {q}Gq\rangle$, it contains
 two parts: one part could be written as $\rho$, and the other part couldn't,
 which we perform the Borel transformation directly. Therefore
\begin{equation}
\Pi^{\langle \bar{q}Gq\rangle \langle
\bar{q}q\rangle}(M_B^2)=\int_{4m^2}^{\infty} ds
e^{-s/M_B^2}\rho^{\langle \bar{q}Gq\rangle \langle
\bar{q}q\rangle1}(s)+\Pi^{\langle \bar{q}Gq\rangle \langle
\bar{q}q\rangle2}(M_B^2).
\end{equation}

For the interpolating current $\eta_1$:
\begin{equation}
\begin{split}
\rho^{pert}_1(s)&=\frac{384}{\pi^4}\Big[\big(16\rho^L_{115}(s)+m^2 \rho^L_{114}(s)-2 m^2 \rho^K_{114}(s)+6m^2\rho^I_{114}(s)-\rho^O_{114}(s)+2\rho^N_{114}(s)-4\rho^J_{114}(s)\big)\\
&+\frac{\alpha}{\pi}\big(\frac{17}{6}\rho^L_{115}(s)-\frac{25}{24}m^2
\rho^I_{114}(s)\big)\Big]
\end{split}
\end{equation}
\begin{equation}
\begin{split}
\rho^{\langle g^2GG\rangle}_1(s)&=\frac{\langle g^2
GG\rangle}{\pi^4}\Big[
5\rho^J_{123}(s)+\frac{23}{8}\rho^N_{123}(s)-\frac{2}{3}\rho^N_{213}(s)-\frac{10}{3}\rho^N_{224}(s)+\frac{1}{6}\rho^O_{123}(s)+\frac{1}{3}\rho^O_{213}(s)
-\frac{10}{3}\rho^O_{224}(s)
\\
&-\frac{39}{8}\rho^I_{123}(s)+1152m^2\rho^I_{134}(s)-20 m^2 \rho^I_{224}(s)-16 m^2 \rho^J_{133}(s)-768 m^2 \rho^J_{144}(s)-\frac{5}{3}\rho^J_{223}(s)\\
&+192 m^2\big( \rho^N_{144}(s)+\rho^N_{414}(s)\big)-96 m^2 \big( \rho^O_{144}(s)+\rho^O_{414}(s)\big)+1152 m^4 \rho^I_{144}(s)-\frac{95}{48}\rho^K_{113}(s)+\frac{5}{6}m^2 \rho^K_{123}(s)\\
& -192m^4\big(\rho^K_{144}(s)+\rho^K_{414}(s)\big)+\frac{10}{3} m^2 \rho^K_{213}(s)-\frac{5}{3}\rho^K_{214}(s)-\frac{5}{3}m^2\rho^K_{224}(s)-240m^2\rho^K_{314}(s)+\frac{91}{48}\rho^L_{113}(s)\\
&-\frac{m^2}{2}\rho^L_{123}(s)-\frac{29}{3}\rho^L_{124}(s)+160m^2\rho^L_{134}(s)+3072m^2\rho^L_{145}(s)+10m^2\rho^L_{224}(s)
-\frac{31}{12}\rho^O_{124}(s)-\frac{8}{3}\rho^O_{214}(s)\\
&-192m^2\rho^K_{134}(s)+192 m^4\rho^L_{144}(s) \Big]
\end{split}
\end{equation}

\begin{equation}
\begin{split}
\rho^{\langle \bar{q}Gq\rangle}_1(s)=-&\frac{m \langle
\bar{q}g\sigma\cdot G q\rangle}{3\pi^2}\big( 29 \rho^I_{112}(s)+39m^2\rho^I_{122}(s)-5
\rho^N_{212}(s)-116 \rho^M_{123}(s)-40 \rho^M_{213}(s)+10m^2
\rho^K_{122}(s)+5m^2 \rho^{K}_{212}(s)\\
&+\frac{521}{8}  \rho^K_{112}(s)\big)
\end{split}
\end{equation}

\begin{align}
\rho&^{\langle\bar{q}q\rangle^2}_1(s)=\frac{16}{3}\langle\bar{q}q\rangle^2 \big( \rho^Q_{110}(s)-m^4\rho^I_{110}(s)\big)  \\
\rho&^{\langle\bar{q}q\rangle\langle \bar{q}Gq\rangle1}_1(s)=\langle\bar{q}q\rangle\langle \bar{q}Gq\rangle \Big(\frac{1}{18}\rho^Q_{120}(s)-\frac{119}{36}m^2\rho^I_{110}(s)-\frac{1}{18} m^4\rho^I_{120}(s)-\frac{2}{3}\rho^P_{110}(s)+\frac{61 }{48}\rho^N_{110}(s)+\frac{61}{48}m^2\rho^K_{110}(s)\Big)  \\
\Pi&^{\langle\bar{q}q\rangle\langle
\bar{q}Gq\rangle2}_1(M_B^2)=\frac{2}{3}\langle\bar{q}q\rangle\langle
\bar{q}Gq\rangle \big(m^4\Pi^I(M_B^2)-\Pi^{II}(M_B^2)\big)
\end{align}

%
%

For the interpolating current $\eta_2$:
\begin{equation}
\begin{split}
\rho^{pert}_2(s)&=\frac{192}{\pi^4}\Big[\big(16\rho^L_{115}(s)+m^2 \rho^L_{114}(s)-2 m^2 \rho^K_{114}(s)+6m^2\rho^I_{114}(s)-\rho^O_{114}(s)+2\rho^N_{114}(s)-4\rho^J_{114}(s)\big)\\
&+\frac{\alpha}{\pi}\big(\frac{7}{6}\rho^L_{115}(s)-\frac{5}{24}m^2\rho^I_{114}(s)\big)\Big]
\end{split}
\end{equation}
\begin{equation}
\begin{split}
\rho^{\langle g^2GG\rangle}_2(s)&=\frac{\langle g^2
GG\rangle}{\pi^4}\Big[
-3\rho^J_{123}(s)+\frac{7}{8}\rho^N_{123}(s)+\frac{2}{3}\rho^N_{213}(s)-\frac{2}{3}\rho^N_{224}(s)-\frac{7}{6}\rho^O_{123}(s)
-\frac{1}{3}\rho^O_{213}(s)-\frac{1}{3}\rho^O_{224}(s)\\
&+\frac{57}{8}m^2\rho^I_{123}(s)+576m^2\rho^I_{134}(s)-4m^2\rho^I_{224}(s)-8m^2\rho^J_{133}(s)-384m^2\rho^J_{144}(s)-\frac{1}{3}m^2\rho^J_{223}(s)\\
&+96m^2\big(\rho^N_{144}(s)+\rho^N_{414}(s)\big)-48m^2\big(\rho^O_{144}(s)+\rho^O_{144}(s)\big)+576m^4\rho^I_{144}(s)+\frac{17}{48}\rho^K_{113}(s)+\frac{1}{6} m^2 \rho^K_{123}(s)\\
&-96 m^2 \rho^K_{134}(s)-96 m^4 \big(\rho^K_{144}(s)+\rho^K_{414}(s)\big)-\frac{10}{3} m^2 \rho^K_{213}(s)-\frac{1}{3}\rho^K_{214}(s)-\frac{1}{3} m^2 \rho^K_{224}(s)\\
&-120 m^2 \rho^K_{314}(s)+\frac{47}{48}\rho^L_{113}(s)+\frac{3}{2}m^2 \rho^L_{123}(s)+\frac{35}{3}\rho^L_{124}(s)+80m^2 \rho^L_{134}(s)+96 m^4 \rho^L_{144}(s)\\&+1536 m^2 \rho^L_{145}(s)+2 m^2 \rho^L_{224}(s)+\frac{1}{12}\rho^H_{124}(s)+\frac{8}{3} \rho^H_{214}(s)\Big]
\end{split}
\end{equation}

\begin{equation}
\begin{split}
\rho^{\langle \bar{q}Gq\rangle}_2(s)=&\frac{m \langle
\bar{q}g\sigma\cdot G q\rangle}{3\pi^2}\big( 11\rho^I_{112}(s)+\rho^N_{212}(s)-44\rho^M_{123}(s)+8\rho^M_{213}(s)+9m^2\rho^I_{122}(s)-\frac{73}{8}\rho^K_{112}(s)+10m^2\rho^K_{122}(s)\\
&-m^2\rho^K_{212}(s)\big)
\end{split}
\end{equation}

\begin{align}
\rho&^{\langle\bar{q}q\rangle^2}_2(s)=\frac{8}{3}\langle\bar{q}q\rangle^2 \big( \rho^Q_{110}(s)-m^4\rho^I_{110}(s)\big)\\
\rho&^{\langle\bar{q}q\rangle\langle
\bar{q}Gq\rangle1}_2(s)=\langle\bar{q}q\rangle\langle
\bar{q}Gq\rangle
\big(\frac{17}{18}\rho^Q_{120}(s)-\frac{79}{36}m^2\rho^I_{110}(s)-\frac{17}{18} m^4\rho^I_{120}(s)-\frac{1}{3}\rho^P_{110}(s)+\frac{29 }{48}\rho^N_{110}(s)+\frac{29}{48}m^2\rho^K_{110}(s)\Big) \\
\Pi&^{\langle\bar{q}q\rangle\langle
\bar{q}Gq\rangle2}_2(M_B^2)=\frac{1}{3}\langle\bar{q}q\rangle\langle
\bar{q}Gq\rangle \big(m^4\Pi^I(M_B^2)-\Pi^{II}(M_B^2)\big)
\end{align}

For the interpolating current $\eta_3$:
\begin{equation}
\begin{split}
\rho^{pert}_3(s)&=\frac{384}{\pi^4}\Big[\big(16\rho^L_{115}(s)+m^2 \rho^L_{114}(s)-2 m^2 \rho^K_{114}(s)+6m^2\rho^I_{114}(s)-\rho^O_{114}(s)+2\rho^N_{114}(s)-4\rho^J_{114}(s)\big)\\
&+\frac{\alpha}{\pi}\big(\frac{17}{6}\rho^L_{115}(s)-\frac{25}{24}m^2
\rho^I_{114}(s)\big)\Big]
\end{split}
\end{equation}

\begin{equation}
\begin{split}
\rho^{\langle g^2GG\rangle}_3(s)&=\frac{\langle g^2
GG\rangle}{\pi^4}\Big[
5\rho^J_{123}(s)+\frac{23}{8}\rho^N_{123}(s)-\frac{2}{3}\rho^N_{213}(s)-\frac{10}{3}\rho^N_{224}(s)+\frac{1}{6}\rho^O_{123}(s)+\frac{1}{3}\rho^O_{213}(s)
-\frac{10}{3}\rho^O_{224}(s)
\\
&-\frac{39}{8}\rho^I_{123}(s)+1152m^2\rho^I_{134}(s)-20 m^2 \rho^I_{224}(s)-16 m^2 \rho^J_{133}(s)-768 m^2 \rho^J_{144}(s)-\frac{5}{3}\rho^J_{223}(s)\\
&+192 m^2\big( \rho^N_{144}(s)+\rho^N_{414}(s)\big)-96 m^2 \big( \rho^O_{144}(s)+\rho^O_{414}(s)\big)+1152 m^4 \rho^I_{144}(s)-\frac{95}{48}\rho^K_{113}(s)+\frac{5}{6}m^2 \rho^K_{123}(s)\\
& -192m^4\big(\rho^K_{144}(s)+\rho^K_{414}(s)\big)+\frac{10}{3} m^2 \rho^K_{213}(s)-\frac{5}{3}\rho^K_{214}(s)-\frac{5}{3}m^2\rho^K_{224}(s)-240m^2\rho^K_{314}(s)+\frac{91}{48}\rho^L_{113}(s)\\
&-\frac{m^2}{2}\rho^L_{123}(s)-\frac{29}{3}\rho^L_{124}(s)+160m^2\rho^L_{134}(s)+3072m^2\rho^L_{145}(s)+10m^2\rho^L_{224}(s)
-\frac{31}{12}\rho^O_{124}(s)-\frac{8}{3}\rho^O_{214}(s)\\
&-192m^2\rho^K_{134}(s)+192 m^4\rho^L_{144}(s) \Big]
\end{split}
\end{equation}

\begin{equation}
\begin{split}
\rho^{\langle \bar{q}Gq\rangle}_3(s)=&\frac{m \langle
\bar{q}g\sigma\cdot G q\rangle}{3\pi^2}\big( 29 \rho^I_{112}(s)+39m^2\rho^I_{122}(s)-5
\rho^N_{212}(s)-116 \rho^M_{123}(s)-40 \rho^M_{213}(s)+10m^2
\rho^K_{122}(s)+5m^2 \rho^{K}_{212}(s)\\
&+\frac{521}{8}  \rho^K_{112}(s)\big)
\end{split}
\end{equation}

\begin{align}
\rho&^{\langle\bar{q}q\rangle^2}_3(s)=\frac{16}{3}\langle\bar{q}q\rangle^2 \big( \rho^Q_{110}(s)-m^4\rho^I_{110}(s)\big)  \\
\rho&^{\langle\bar{q}q\rangle\langle \bar{q}Gq\rangle1}_3(s)=\langle\bar{q}q\rangle\langle \bar{q}Gq\rangle \Big(\frac{1}{18}\rho^Q_{120}(s)-\frac{119}{36}m^2\rho^I_{110}(s)-\frac{1}{18} m^4\rho^I_{120}(s)-\frac{2}{3}\rho^P_{110}(s)+\frac{61 }{48}\rho^N_{110}(s)+\frac{61}{48}m^2\rho^K_{110}(s)\Big)  \\
\Pi&^{\langle\bar{q}q\rangle\langle
\bar{q}Gq\rangle2}_3(M_B^2)=\frac{2}{3}\langle\bar{q}q\rangle\langle
\bar{q}Gq\rangle \big(m^4\Pi^I(M_B^2)-\Pi^{II}(M_B^2)\big)
\end{align}


For the interpolating current $\eta_4$:
\begin{equation}
\begin{split}
\rho^{pert}_4(s)&=\frac{192}{\pi^4}\Big[\big(16\rho^L_{115}(s)+m^2 \rho^L_{114}(s)-2 m^2 \rho^K_{114}(s)+6m^2\rho^I_{114}(s)-\rho^O_{114}(s)+2\rho^N_{114}(s)-4\rho^J_{114}(s)\big)\\
&+\frac{\alpha}{\pi}\big(\frac{7}{6}\rho^L_{115}(s)-\frac{5}{24}m^2\rho^I_{114}(s)\big)\Big]
\end{split}
\end{equation}
\begin{equation}
\begin{split}
\rho^{\langle g^2GG\rangle}_4(s)&=\frac{\langle g^2
GG\rangle}{\pi^4}\Big[
-3\rho^J_{123}(s)+\frac{7}{8}\rho^N_{123}(s)+\frac{2}{3}\rho^N_{213}(s)-\frac{2}{3}\rho^N_{224}(s)-\frac{7}{6}\rho^O_{123}(s)
-\frac{1}{3}\rho^O_{213}(s)-\frac{1}{3}\rho^O_{224}(s)\\
&+\frac{57}{8}m^2\rho^I_{123}(s)+576m^2\rho^I_{134}(s)-4m^2\rho^I_{224}(s)-8m^2\rho^J_{133}(s)-384m^2\rho^J_{144}(s)-\frac{1}{3}m^2\rho^J_{223}(s)\\
&+96m^2\big(\rho^N_{144}(s)+\rho^N_{414}(s)\big)-48m^2\big(\rho^O_{144}(s)+\rho^O_{144}(s)\big)+576m^4\rho^I_{144}(s)+\frac{17}{48}\rho^K_{113}(s)+\frac{1}{6} m^2 \rho^K_{123}(s)\\
&-96 m^2 \rho^K_{134}(s)-96 m^4 \big(\rho^K_{144}(s)+\rho^K_{414}(s)\big)-\frac{10}{3} m^2 \rho^K_{213}(s)-\frac{1}{3}\rho^K_{214}(s)-\frac{1}{3} m^2 \rho^K_{224}(s)\\
&-120 m^2 \rho^K_{314}(s)+\frac{47}{48}\rho^L_{113}(s)+\frac{3}{2}m^2 \rho^L_{123}(s)+\frac{35}{3}\rho^L_{124}(s)+80m^2 \rho^L_{134}(s)+96 m^4 \rho^L_{144}(s)\\&+1536 m^2 \rho^L_{145}(s)+2 m^2 \rho^L_{224}(s)+\frac{1}{12}\rho^H_{124}(s)+\frac{8}{3} \rho^H_{214}(s)\Big]
\end{split}
\end{equation}
\begin{equation}
\begin{split}
\rho^{\langle \bar{q}Gq\rangle}_4(s)=-&\frac{m \langle
\bar{q}g\sigma\cdot G q\rangle}{3\pi^2}\big( 11\rho^I_{112}(s)+\rho^N_{212}(s)-44\rho^M_{123}(s)+8\rho^M_{213}(s)+9m^2\rho^I_{122}(s)-\frac{73}{8}\rho^K_{112}(s)+10m^2\rho^K_{122}(s)\\
&-m^2\rho^K_{212}(s)\big)
\end{split}
\end{equation}

\begin{align}
\rho&^{\langle\bar{q}q\rangle^2}_4(s)=\frac{8}{3}\langle\bar{q}q\rangle^2 \big( \rho^Q_{110}(s)-m^4\rho^I_{110}(s)\big)\\
\rho&^{\langle\bar{q}q\rangle\langle
\bar{q}Gq\rangle1}_4(s)=\langle\bar{q}q\rangle\langle
\bar{q}Gq\rangle
\big(\frac{17}{18}\rho^Q_{120}(s)-\frac{79}{36}m^2\rho^I_{110}(s)-\frac{17}{18} m^4\rho^I_{120}(s)-\frac{1}{3}\rho^P_{110}(s)+\frac{29 }{48}\rho^N_{110}(s)+\frac{29}{48}m^2\rho^K_{110}(s)\Big) \\
\Pi&^{\langle\bar{q}q\rangle\langle
\bar{q}Gq\rangle2}_4(M_B^2)=\frac{1}{3}\langle\bar{q}q\rangle\langle
\bar{q}Gq\rangle \big(m^4\Pi^I(M_B^2)-\Pi^{II}(M_B^2)\big)
\end{align}

The functions $\rho_{hjk}^{I,J,K \dots}(s) $ and
$\Pi^{I,II}(M_B^2)$ in the above expressions are defined as:

\begin{equation}
\rho^I_{hjk}(s)=\frac{ (-1)^{k} 4^{-k-2}}{ \pi^2 \Gamma (h) \Gamma
(j) \Gamma (k)\Gamma
(3-h-j+k)}\int^{\alpha_{max}}_{\alpha_{min}}d\alpha
\int^{\beta_{max}}_{\beta_{min}}d\beta \frac{(\alpha +\beta
-1)^{k-1} \left(m^2 (\alpha +\beta )-\alpha  \beta
s\right)^{2-h-j+k}}{\alpha ^{1+k-h} \beta ^{1+k-j} },
\end{equation}
\begin{equation}
\begin{split}
\rho^J_{hjk}&(s)=\frac{ (-1)^{k}  4^{-k-2} }{\pi^2  \Gamma (h) \Gamma (j) \Gamma (k) \Gamma (4-h-j+k)}\\
&\int^{\alpha_{max}}_{\alpha_{min}}d\alpha
\int^{\beta_{max}}_{\beta_{min}}d\beta  \frac{(\alpha +\beta
-1)^{k-1} \left(m^2 (\alpha +\beta )-\alpha  \beta
s\right)^{2-h-j+k} \left(2 m^2 (\alpha +\beta )+\alpha  \beta
s(h+j-k-5)\right)}{\alpha^{1+k-h} ~\beta^{1+k-j}},
\end{split}
\end{equation}
where $h,j,k>0$, $h+j-k\leq 2$.
\begin{equation}
\begin{split}
\rho^K_{hjk}&(s)=\frac{(-1)^{k} 2^{-2 k-3}}{\pi^2  \Gamma (h)
\Gamma (j) \Gamma (k) \Gamma (3-h-j+k)} \\
&\int^{\alpha_{max}}_{\alpha_{min}}d\alpha
\int^{\beta_{max}}_{\beta_{min}}d\beta  \frac{ (\alpha +\beta
-1)^{k-1} \left(m^2 (\alpha +\beta )-\alpha  \beta
s\right)^{1-h-j+k} \left(2 m^2 (\alpha +\beta )+\alpha  \beta  s
(h+j-k-4)\right)}{\alpha ^{1+k-h} \beta ^{k-j}},
\end{split}
\end{equation}
where $h,j,k>0$, and $h+j-k\leq 1$.
\begin{equation}
\begin{split}
\rho^L_{hjk}(s)&=\frac{ (-1)^{k} 4^{-k-1}}{\pi^2  \Gamma (h) \Gamma (j) \Gamma (k) \Gamma (3-h-j+k)} \int^{\alpha_{max}}_{\alpha_{min}}d\alpha \int^{\beta_{max}}_{\beta_{min}}d\beta \frac{(\alpha+\beta-1)^{k-1}(m^2(\alpha+\beta)-\alpha \beta s)^{-h-j+k}}{\alpha^{k-h}\beta^{k-j}} \\
& \big[ 6 (m^2(\alpha +\beta)-\alpha \beta s)^2-\alpha \beta
s\big( 6(m^2(\alpha+\beta)-\alpha\beta s)(2+k-h-j)-\alpha \beta s
(2+k-h-j)(1+k-h-j)\big) \big],
\end{split}
\end{equation}
where, $h,j,k>0$, and $h+j-k\leq 0$.
\begin{equation}
\rho^L_{144}(s)=\frac{3  }{512
\pi^2\Gamma(4)^2}\int^{\alpha_{max}}_{\alpha_{min}}d\alpha
\int^{\beta_{max}}_{\beta_{min}}d\beta
\frac{(\alpha+\beta-1)^3\big(m^2(\alpha+\beta)-2\alpha \beta
s\big)}{\alpha^3}-\frac{
m^4}{\pi^2}\int^{\alpha_{max}}_{\alpha_{min}}d\alpha \frac{\alpha
s^2 \big( m^2-(1-\alpha)\alpha s \big)^3}{36864 (m^2-\alpha s)^6},
\end{equation}
\begin{equation}
\begin{split}
\rho^N_{hjk}&(s)=\frac{ (-1)^{k+1} 4^{-k-2}}{\pi ^2 \Gamma (h) \Gamma (j) \Gamma (k) \Gamma (4-h-j+k)}\int^{\alpha_{max}}_{\alpha_{min}}d\alpha \int^{\beta_{max}}_{\beta_{min}}d\beta \frac{(\alpha +\beta -1)^{k-1} \left(m^2 (\alpha +\beta )-\alpha  \beta  s\right)^{-h-j+k+1}}{\alpha^{k+1-h} \beta^{k+1-j}} \\
&\left(2 (6 \alpha -1) \left(\alpha  \beta  s-m^2 (\alpha +\beta
)\right)^2-\alpha  \beta  s (3-h-j+k) \left(\alpha  \beta  s (2
\alpha  (h+j-k-8)+1)+(12 \alpha -1) m^2 (\alpha +\beta
)\right)\right),
\end{split}
\end{equation}
where $h,j,k>0$, $h+j-k\leq 1$.
\begin{equation}
\begin{split}
\rho^O_{hjk}(s)&=\frac{(-1)^{k+1} 2^{-2 k-3}}{\pi ^2 \Gamma (h) \Gamma (j) \Gamma (k) \Gamma (4-h-j+k)}\int^{\alpha_{max}}_{\alpha_{min}}d\alpha \int^{\beta_{max}}_{\beta_{min}}d\beta \frac{(\alpha +\beta -1)^{k-1} \left(m^2 (\alpha +\beta )-\alpha  \beta  s\right)^{-h-j+k}}{\alpha^{k-h}\beta{k-j+1}} \\
& \big[ 6 (8 \alpha -3) \left(m^2 (\alpha +\beta )-\alpha  \beta  s\right)^3-18 (4 \alpha -1) \alpha \beta s (3-h-j+k) \left(m^2 (\alpha +\beta )-\alpha  \beta  s\right)^2 \\
&+3 (8 \alpha -1) \alpha ^2 \beta^2 s^2 (2-h-j+k) (3-h-j+k) \left(m^2 (\alpha +\beta )-\alpha  \beta  s\right) - \\
&2 \alpha ^4 s^3 (1-h-j+k) (2-h-j+k) (1-h-j+k)\big],
\end{split}
\end{equation}
where  $h,j,k>0$, $h+j-k\leq 1$.
\begin{equation}
\begin{split}
&\rho^O_{144}(s)=-\frac{m^4}{36864 \pi^2}\int^{\alpha_{max}}_{\alpha_{min}}d\alpha \frac{\alpha^3 s^3 \big(m^2-(1-\alpha)\alpha s\big)^3}{(m^2-\alpha s)^6} \\
&-\frac{1}{2048\pi^2
\Gamma(4)^2}\int^{\alpha_{max}}_{\alpha_{min}}d\alpha
\int^{\beta_{max}}_{\beta_{min}}d\beta \frac{ (\alpha +\beta -1)^3
\left((8 \alpha -3) m^4 (\alpha +\beta )^2-4 \alpha  (10 \alpha
-3) \beta  m^2 s (\alpha +\beta )+10 \alpha ^2 (4 \alpha -1) \beta
^2 s^2\right)}{ \alpha ^3 \beta },
\end{split}
\end{equation}
\begin{equation}
\begin{split}
&\rho^O_{414}(s)=\frac{1}{36864~ \pi^2}\int^{\alpha_{max}}_{\alpha_{min}}d\alpha \frac{\alpha^3 s^3 \big(m^2-(1-\alpha)\alpha s\big)^3}{m^2~(m^2-\alpha s)^3} \\
&-\frac{1}{2048\pi^2
\Gamma(4)^2}\int^{\alpha_{max}}_{\alpha_{min}}d\alpha
\int^{\beta_{max}}_{\beta_{min}}d\beta \frac{ (\alpha +\beta -1)^3
\left((8 \alpha -3) m^4 (\alpha +\beta )^2-4 \alpha  (10 \alpha
-3) \beta  m^2 s (\alpha +\beta )+10 \alpha ^2 (4 \alpha -1) \beta
^2 s^2\right)}{  \beta^4 }.
\end{split}
\end{equation}

\begin{equation}
\begin{split}
\rho^H_{hjk}&(s)=\frac{ (-1)^{k+1} 2^{-2 k-3}}{\pi ^2 \Gamma (h) \Gamma (j) \Gamma (k) \Gamma (3-h-j+k)}\int^{\alpha_{max}}_{\alpha_{min}}d\alpha \int^{\beta_{max}}_{\beta_{min}}d\beta \frac{(\alpha +\beta -1)^{k-1} \left(m^2 (\alpha +\beta )-\alpha  \beta  s\right)^{-h-j+k-1}}{\alpha^{k-h} \beta^{k-j} (1-\alpha)^2} \\
&\Big(12 (1-\alpha )^3 (8 \alpha -1) \left(m^2 (\alpha +\beta )-\alpha  \beta  s\right)^3-3 \left(48 \alpha ^3-100 \alpha ^2+56 \alpha -3\right) \alpha  \beta  s (-h-j+k+2) \left(\alpha  \beta  s-m^2 (\alpha +\beta )\right)^2 \\
&-2 (\alpha -1)^2 (24 \alpha -1) \alpha ^2 \beta ^2 s^2 (h+j-k-2) (h+j-k-1) \left(\alpha  \beta  s-m^2 (\alpha +\beta )\right)\\
&+4 (\alpha -1)^2 \alpha ^4 \beta ^3 s^3 (h+j-k-2) (h+j-k-1)
(h+j-k)\Big),
\end{split}
\end{equation}
where  $h,j,k>0$, $h+j-k\leq -1$.

\begin{align}
\rho^I_{110}(s)&=-\frac{1}{16\pi^2}\sqrt{1-\frac{4m^2}{s}},~~~\\
\rho^I_{120}(s)&=\frac{1}{16\pi^2}\frac{1}{\sqrt{s(s-4m^2)}} ,\\
\rho^Q_{110}(s)&=-\frac{1}{16\pi^2}\Big(\frac{s}{2}-m^2\Big)\sqrt{1-\frac{4m^2}{s}},\\
\rho^Q_{120}(s)&=-\frac{1}{16\pi^2}\Big(\frac{s}{2}-m^2\Big)^2\sqrt{1-\frac{4m^2}{s}},\\
\rho^K_{110}(s)&=-\frac{1}{8\pi^2}\big(\frac{1}{4(1-s/4m^2)}+\sqrt{1-\frac{4m^2}{s}}\big),\\
\rho^N_{110}(s)&=-\frac{2}{32\pi^2}\big( (3s-2m^2)\sqrt{1-\frac{4m^2}{s}}-\frac{4m^4}{\sqrt{s(s-4m^2)}}\big),\\
\rho^P_{110}(s)&=\frac{-83 m^6-559 m^4 s+326 m^2 s^2-53 s^3}{120 \pi ^2 s \sqrt{s \left(s-4 m^2\right)}},\\
\Pi^I(M_B^2)&=\frac{1}{4\pi^2} \int^1_0 dx \frac{m^2}{x^2 M_B^2}\exp{\big[-\frac{m^2}{x(1-x)M_B^2}\big]} ,\\
\Pi^{II}(M_B^2)&=-\frac{1}{8\pi^2}\int^1_0 dx \frac{m^6}{x (1-x)
M_B^2}\exp{\big[-\frac{m^2}{x(1-x)M_B^2}\big]}.
\end{align}

The integration limits appears in the above expression are:

\begin{align}
\qquad \alpha_{max}= & \frac{1+\sqrt{1-4m^2/s}}{2} & \alpha_{min}= & \frac{1-\sqrt{1-4m^2/s}}{2}\qquad \\
\qquad\beta_{max}= & 1-\alpha & \beta_{min}= & \frac{\alpha m^2}{\alpha s-m^2}.\qquad
\end{align}

\end{document}